\documentclass[superscriptaddress,prl,twocolumn,nofootinbib,showpacs,preprintnumbers]{revtex4}

\usepackage{amsfonts}
\usepackage{mathrsfs}
\usepackage{amssymb}
\usepackage{amsmath}
\usepackage{graphicx,epsfig}

\def\bwt{\begin{widetext}}

\def\ewt{\end{widetext}}

\def\be{\begin{equation}}

\def\ee{\end{equation}}

\def\bea{\begin{eqnarray}}

\def\eea{\end{eqnarray}}

\def\bean{\begin{eqnarray*}}

\def\eean{\end{eqnarray*}}

\def\bary{\begin{array}}

\def\eary{\end{array}}

\def\bit{\begin{itemize}}

\def\eit{\end{itemize}}

\def\su5u1{SU(5) \times U(1)}

\def\fsu5u1{SU(5) \times U(1)'}

\def\so10{SO(10)}

\def\sq20{SO(10) \times SO(10)}

\usepackage{amssymb}

\begin{document}

\title{${\cal F}$-$SU(5)$}

\author{Jing Jiang}

\affiliation{Department of Physics, University of Wisconsin, 
Madison, WI 53706, USA}

\author{Tianjun Li}

\affiliation{George P. and Cynthia W. Mitchell Institute for
Fundamental Physics, Texas A$\&$M University, College Station, TX
77843, USA }

\affiliation{ Institute of Theoretical Physics, Chinese Academy of
Sciences, Beijing 100080, China}

\author{Dimitri V. Nanopoulos}

\affiliation{George P. and Cynthia W. Mitchell Institute for
Fundamental Physics,
 Texas A$\&$M University, College Station, TX 77843, USA }

\affiliation{Astroparticle Physics Group,
Houston Advanced Research Center (HARC),
Mitchell Campus, Woodlands, TX 77381, USA}

\affiliation{Academy of Athens, Division of Natural Sciences,
 28 Panepistimiou Avenue, Athens 10679, Greece }

\author{Dan Xie}

\affiliation{George P. and Cynthia W. Mitchell Institute for
Fundamental Physics, Texas A$\&$M University, College Station, TX
77843, USA }



\begin{abstract}

We construct three flipped $SU(5)\times U(1)_X$ models from
F-theory, and consider two such models from free fermionic
string model building. To achieve the decoupling scenario
in F-theory models and the string-scale gauge coupling
unification in free fermionic models, we introduce
vector-like particles at the TeV scale that can
be observed at the Large Hadron Collider. We study
gauge coupling unification, and find that
proton decay is within the reach of the future Hyper-Kamiokande 
experiment. In these models, the doublet-triplet splitting
problem and monopole problem can be solved, the neutrino masses 
and mixings can be explained via the double seesaw or seesaw mechanism, 
the observed baryon asymmetry can be obtained through leptogenesis, 
the hybrid inflation can be realized, and
the correct cosmic primodial density fluctuations can be generated.

\end{abstract}


\preprint{ACT-06-08, MIFP-08-30}

\maketitle


{\bf Introduction~--}~The goal of string phenomenology is to 
construct realistic string models with moduli stabilization 
and without chiral exotic particles, and then make clean predictions 
that can be tested at the Large Hadron Collider (LHC) and other
experiments. Previously, string model building has been studied 
extensively in the heterotic $E_8\times E_8$ string theory and 
Type II string theories with D-branes.

Recently,  Grand Unified Theories (GUTs) have been constructed 
in the F-theory which can be considered as the
strongly coupled formulation of ten-dimensional Type IIB string 
theory with a varying axion ($a$)-dilaton ($\phi$) field 
$\tau=a+ie^{-\phi}$~\cite{Vafa:1996xn, Beasley:2008dc, Donagi:2008ca}.
The gauge fields are on the observable seven-branes that wrap a del Pezzo 
$n$ ($dP_n$) surface for the extra four space dimensions, 
while the matter and Higgs fields are localized on the complex
codimension one curves (two-dimensional subspaces) in $dP_n$. 
Because the GUT scale $M_{\rm GUT}$ is about $2.4\times 10^{16}$ GeV 
while the Planck scale $M_{\rm Pl}$ is about $10^{19}$ GeV,
$M_{\rm GUT}/M_{\rm Pl}$ is a small number around $10^{-3}-10^{-2}$.
Also,  the known GUTs without additional 
chiral exotic particles are asymptotically free. Thus, 
we can consider the decoupling scenario where $M_{\rm GUT}/M_{\rm Pl}$ 
is small and then the gravity can be decoupled. 
In the decoupling limit where 
$M_{\rm Pl} \rightarrow \infty$ while $M_{\rm GUT}$ remains 
finite, semi-realistic $SU(5)$ models and $SO(10)$ models 
without chiral exotic particles have been constructed 
locally~\cite{Beasley:2008dc, Donagi:2008ca}.
A brand new feature is that 
the $SU(5)$ and $SO(10)$ gauge symmetries can be broken down 
to the Standard Model (SM) and $SU(5)\times U(1)$ gauge symmetries 
via $U(1)$ fluxes, respectively. It seems to us that 
$SO(10)$ models are more attractive than $SU(5)$ models since they 
have SM fermion unification. In $SO(10)$ models, in order 
to eliminate the zero modes of chiral exotic particles,  
we must break the $SO(10)$ gauge symmetry down to the 
flipped $SU(5)\times U(1)_X$ gauge 
symmetry~\cite{Beasley:2008dc}.
Interestingly, in flipped $SU(5)\times U(1)_X$ 
models~\cite{smbarr, dimitri}, 
we can solve the doublet-triplet splitting problem via 
the missing partner mechanism~\cite{AEHN-0}.

In the flipped $SU(5)\times U(1)_X$ models with
$SO(10)$ origin, we can have two unification
scales: the $SU(2)_L \times SU(3)_C$ unification scale $M_{23}$
and the $SU(5)\times U(1)_X$ unfication scale $M_U$
where $M_{23}$ is about the usual GUT scale  around
$2\times 10^{16}$ GeV. Thus, with additional 
vector-like particles, we can 
solve the little hierarchy problem between  
the GUT scale and string scale $M_{\rm string}$ in
the weakly coupled heterotic string models where 
\begin{eqnarray}
M_{\rm string} = g_{\rm string} \times 5.27 \times 10^{17} ~{\rm GeV}
\sim 5 \times 10^{17} ~{\rm GeV}~,~\,
\label{St-Unif}
\end{eqnarray}
where $g_{\rm string}$ is the string coupling constant and around
order 1 (${\cal O} (1)$)~\cite{Lopez:1995cs, Jiang:2006hf}. 
Similarly, for flipped $SU(5)\times U(1)_X$ models
from F-theory, we can naturally realize the decoupling scenario 
where $M_{23}/M_U$ or  $M_{23}/M_{\rm Pl}$ can be small.

In this paper, we construct three flipped $SU(5)\times U(1)$ 
models from F-theory, and consider
two such models derived from the four-dimensional free fermionic 
formulation of the weakly coupled heterotic string theory~\cite{LNY}.
To achieve the decoupling scenario in F-theory models
and the string-scale gauge coupling unification in 
free fermionic models, we introduce vector-like particles which 
form complete flipped $SU(5)\times U(1)_X$ multiplets. 
In order to avoid the Landau pole problem for 
the strong coupling constant, we can introduce 
{\it two and only two} 
sets of vector-like particles at the TeV scale.
Because there is one set of such TeV-scale 
vector-like particles for
each flipped $SU(5)\times U(1)_X$ model,  
our models can definitely be tested at the LHC. 
We call these flipped $SU(5)\times U(1)$ models
as ${\cal F}$-$SU(5)$ where ${\cal F}$ means not only 
``flipped'' but also ``F''-theory or ``free 
fermionic'' constructions. 
We study the gauge coupling unification at two loops,
and show that proton decay
is within the reach of the future Hyper-Kamiokande 
experiment~\cite{Nakamura:2003hk}
since the $SU(2)_L\times SU(3)_C$ unified gauge coupling
is stronger than that in the traditional flipped 
$SU(5)\times U(1)_X$ models due to the TeV-scale
vector-like particles.
In these models, we can explain the neutrino masses and mixings
via double seesaw or seesaw mechanism, obtain the observed 
baryon asymmetry by leptogenesis, realize hybrid inflation,
solve the monopole problem, 
and generate the correct cosmic primodial density fluctuations.
In this paper, we shall only present three models from
F-theory as examples, and the detailed study on F-theory 
model building and phenomenological consequences will 
be presented elsewhere~\cite{JLNX-P}.

{\bf Flipped $SU(5)\times U(1)_X$ Models~--}~We first 
briefly review the flipped
$SU(5)\times U(1)_X$ models~\cite{smbarr, dimitri, AEHN-0}. 
There are three families of the SM fermions 
whose quantum numbers under $SU(5)\times U(1)_{X}$ are
\bea
F_i={\mathbf{(10, 1)}},~ {\bar f}_i={\mathbf{(\bar 5, -3)}},~
{\bar l}_i={\mathbf{(1, 5)}},
\label{smfermions}
\eea
where $i=1, 2, 3$. In terms of the SM fermions we have
\bea
F_i=(Q_i, D^c_i, N^c_i),~{\overline f}_i=(U^c_i, L_i),
~{\overline l}_i=E^c_i~,~
\label{smparticles}
\eea
where $Q_i$ and $L_i$ are respectively the left-handed
quark and lepton doublets, and $U_i^c$, $D_i^c$, 
$E_i^c$ and $N_i^c$ are
the right-handed up-type quarks,
down-type quarks, leptons and neutrinos, respectively.

To break the GUT and electroweak gauge symmetries, we 
introduce two pairs of Higgs fields
\bea
H={\mathbf{(10, 1)}},~{\overline{H}}={\mathbf{({\overline{10}}, -1)}},
~h={\mathbf{(5, -2)}},~{\overline h}={\mathbf{({\bar {5}}, 2)}},
\label{Higgse1}
\eea
where the particle assignments of Higgs fields are 
\bea
H=(Q_H, D_H^c, N_H^c)~,~
{\overline{H}}= ({\overline{Q}}_{\overline{H}}, {\overline{D}}^c_{\overline{H}}, 
{\overline {N}}^c_{\overline H})~,~\,
\label{Higgse2}
\eea
\bea
h=(D_h, D_h, D_h, H_d)~,~
{\overline h}=({\overline {D}}_{\overline h}, {\overline {D}}_{\overline h},
{\overline {D}}_{\overline h}, H_u)~,~\,
\label{Higgse3}
\eea
where $H_d$ and $H_u$ are one pair of Higgs doublets in the supersymmetric SM.
We also add a SM singlet field $\Phi$.

To break the $SU(5)\times U(1)_{X}$ gauge symmetry down to the SM
gauge symmetry, we introduce the following Higgs superpotential at the GUT scale
\bea
{\it W}_{\rm GUT}=\lambda_1 H H h + \lambda_2 {\overline H} {\overline H} {\overline
h} + \Phi ({\overline H} H-M_{\rm H}^2). 
\label{spgut} 
\eea 
There is only
one F-flat and D-flat direction, which can always be rotated along
the $N^c_H$ and ${\overline {N}}^c_{\overline H}$ directions. So, we obtain that
$<N^c_H>=<{\overline {N}}^c_{\overline H}>=M_{\rm H}$. In addition, the
superfields $H$ and ${\overline H}$ are eaten and acquire large masses via
the supersymmetric Higgs mechanism, except for $D_H^c$ and 
${\overline {D}}^c_{\overline H}$. The superpotential term 
$ \lambda_1 H H h$ and
$ \lambda_2 {\overline H} {\overline H} {\overline h}$ couple the $D_H^c$ and
${\overline {D}}^c_{\overline H}$ with the $D_h$ and ${\overline {D}}_{\overline h}$,
respectively, to form the massive eigenstates with masses
$2 \lambda_1 <N_H^c>$ and $2 \lambda_2 <{\overline {N}}^c_{\overline H}>$. So, we
naturally have the doublet-triplet splitting due to the missing
partner mechanism~\cite{AEHN-0}. 
Because the triplets in $h$ and ${\overline h}$ only have
small mixing through the $\mu h {\overline h}$ term, 
we solve the dimension-5 
proton decay problem from colored Higgsino exchange.

The Yukawa superpotential is
\bea 
{ W}_{\rm Yukawa} &=& y_{ij}^{D}
F_i F_j h + y_{ij}^{U \nu} F_i  {\overline f}_j {\overline
h}+ y_{ij}^{E} {\overline l}_i  {\overline f}_j h + \phi h {\overline h}
\nonumber \\&&
+ y_{ij}^{N} \phi_i {\overline H} F_j 
+ y_{ijk}^{\phi} \phi_i \phi_j \phi_k ~,~\,
\label{potgut}
\eea
where $\phi$ and $\phi_i$ with $i=1, 2, 3$ are SM singlet fields,
$y_{ij}^{D}$, $y_{ij}^{U \nu}$, $y_{ij}^{E}$ and $y_{ij}^{N}$
are Yukawa couplings. The $\mu h {\overline h}$ term is generated 
after $\phi$ obtains a vacuum expectation value.

To separate the scales $M_{23}$ and $M_U$
and then obtain the decoupling scenario in F-theory models
or the string-scale gauge coupling unification in 
free fermionic models,
we introduce the vector-like particles which form complete 
flipped $SU(5)\times U(1)_X$ multiplets. 
In order to avoid the Landau pole
problem for the strong coupling constant, we can only introduce the
following two sets of vector-like particles  around the TeV 
scale~\cite{Jiang:2006hf}
\begin{eqnarray}
&& Z0:  XF ={\mathbf{(10, 1)}}~,~
{\overline{XF}}={\mathbf{({\overline{10}}, -1)}}~;~\\
&& Z1: XF~,~{\overline{XF}}~,~Xl={\mathbf{(1, -5)}}~,~
{\overline{Xl}}={\mathbf{(1, 5)}}
~.~\,
\end{eqnarray}

{\bf ${\cal F}$-$SU(5)$ from F-Theory Model Building~--}~In 
F-theory model building~\cite{Beasley:2008dc, Donagi:2008ca}, 
we introduce the seven-branes which 
wrap a four-dimensional internal subspace of the six compact extra 
dimensions. In the internal geometry, these seven-branes generically 
form intersections over two-dimensional Riemann surfaces, and 
triple intersections at points as well. Near the seven-branes,
the  profiles of axio-dilaton $\tau$ become singular and 
develop  singularities 
with $ADE$ Lie groups $SU(N)$, $SO(2N)$, and $E_6$, 
$E_7$ and $E_8$. These $ADE$ singularities give the corresponding
gauge groups on the seven-branes. On the Riemann surfaces
at the intersections of the seven-branes, the $\tau$ singularities
will be enhanced, and then we can have  matter and Higgs fields
by turning on suitable background fluxes. In addition, at the triple 
intersection points, the $\tau$ singularities will be enhanced further 
and then we can have Yukawa couplings in superpotential.

Since F-theory models are constructed locally, we only concentrate
on the observable seven-branes. To get rid of adjoint chiral
superfields, we consider the del Pezzo 8 ($dP_8$) surface $S$ that
has two complex dimensions, and the observable seven-branes wrap it. 
Because the $dP_8$ surface is contractable, we naturally realize  
the decoupling scenario. Also, the $dP_8$ surface is obtained by
blowing up the projective space $\mathbb{P}^2$ at eight points in general
positions, and its homology is generated by the hyperplane class
$H$ for $\mathbb{P}^2$ and the exceptional classes $E_1$, $E_2$, ..., $E_8$.
We assume that the observable gauge group on $dP_8$ surface $S$ 
is $SO(10)$. On codimension one curves that are  the intersections 
of the observable seven-branes and other
seven-branes, we obtain the
matter fields, Higgs fields, and extra vector-like particles. To break the 
$SO(10)$ gauge symmetry down to
the flipped $SU(5)\times U(1)_X$ gauge symmetry, we turn on 
the $U(1)_X$ flux on $S$ specified by the line bundle $L$. 
To obtain the matter fields, Higgs fields and vector-like particles, 
we also turn on the $U(1)$ fluxes on the seven-branes that intersect
with the observable seven-branes, and we specify these fluxes
 by the line bundle $L^{\prime n}$.

In this paper, we shall only present three models for simplicity,
and the detail model building will be given elsewhere~\cite{JLNX-P}.
We take line bundle $L=\mathcal{O}_{S}(E_{1}-E_{2}+E_4-E_5+E_6-E_7)^{1/4}$.
Note that $\chi(S, L^4)=-2$, we have two pairs of vector-like particles
$XB_i$ and $\overline{XB}_i$ on the bulk $S$, whose $SU(5)\times U(1)_X$
quantum numbers are ${\mathbf{(10, -4)}}$ and 
${\mathbf{({\overline{10}}, 4)}}$, respectively. Moreover,
the curves with homology classes for the matter fields, Higgs fields 
and vector-like particles, and the gauge bundle assignments for 
each curve in our ${\cal F}$-$SU(5)$ models are given 
in Table~\ref{FSU5}. From this table, we obtain:
all the SM fermions are localized on the matter curve
$\Sigma_{F}$; the Higgs fields $h$, $\overline{h}$,
and $(H,~\overline{H})$ are localized on the
curves $\Sigma_{h}$, $\Sigma_{\overline{h}}$, and
$\Sigma_{H}$, respectively; and the vector-like particles
($XF,~\overline{XF}$), ($Xl,~\overline{Xl}$), and
($Xl_{j},~\overline{Xl}_j$) are localized on
the curves $\Sigma_{XF}$, $\Sigma_{Xl} $, 
and $\Sigma^{kl}_{Xl} $, respectively, where 
$j=1, 2,..., 6$ and $kl=15, 17, 42, 47, 62, 65$. 
In addition, there exist singlets
from the intersections of the other seven-branes.
All the curves except $\Sigma_{h}$ and $\Sigma_{\overline{h}}$
are pinched, so, we can realize the trilinear
Yukawa superpotential terms
in Eqs. (\ref{spgut}) and  (\ref{potgut}). The linear
superpotential term  $-M_H^2\Phi$  in Eq. (\ref{spgut})
may be generated via instanton effects.

\begin{table}[htb]
\begin{center}
\begin{tabular}{|c|c|c|c|c|c|}
\hline
 Fields & Curves  & ${\rm Class}$ & $g_{\Sigma}$ &
$L_{\Sigma}$ & $L_{\Sigma}^{\prime n}$\\\hline
$h$ & $\Sigma_{h}$ & $2H-E_{2}-E_{3}$ & $0$ &
$\mathcal{O}(-1)^{1/4}$ & $\mathcal{O}%
(1)^{1/2}$\\\hline
$\overline{h}$ & $\Sigma_{\overline{h}}$ & $2H-E_{1}-E_{3}$ & $0$ &
$\mathcal{O}(1)^{1/4}$ & $\mathcal{O}(-1)^{1/2}$\\\hline
$16_i$ & $\Sigma_{F}$ & $3H$ & $1$ & $\mathcal{O}(0)$ & 
$\mathcal{O}(3p^{\prime})$\\\hline
$H,\overline{H}$ & $\Sigma_{H}$ &
$3H-E_{1}-E_{2}$ & $1$ & $\mathcal{O}(p_{12})^{1/4}$ &
$\mathcal{O}(p_{12})^{-1/4}$\\\hline
$XF,\overline{XF} $ & $\Sigma_{XF}$ &
$3H-E_{4}-E_{5}$ & $1$ & $\mathcal{O}(p_{45})^{1/4}$ &
$\mathcal{O}(p_{45})^{-1/4}$\\\hline
 $Xl,\overline{Xl}$ & $\Sigma_{Xl} $ &
$3H-E_{6}-E_{7}$ & $1$ & $\mathcal{O}(p_{67})^{1/4}$ &
$\mathcal{O}(p_{67})^{-5/4}$\\\hline
$Xl_{j},\overline{Xl}_j$ & $\Sigma^{kl}_{Xl} $ &
$3H-E_{k}-E_{l}$ & $1$ & $\mathcal{O}(p_{kl})^{1/4}$ &
$\mathcal{O}(p_{kl})^{-5/4}$\\\hline
\end{tabular}
\end{center}
\caption{ The particle curves and 
gauge bundle assignments for each curve in the flipped 
$SU(5)\times U(1)_X$ models from F-theory. Here 
$i=1,~2,~3$, $j=1, 2,..., 6$, $kl=15, 17, 42, 47, 62, 65$, and 
$p_{\alpha\beta} =p^{\alpha\beta}_{\alpha}-p^{\alpha\beta}_{\beta}$.}
\label{FSU5}
\end{table}



The vector-like particles $XB_i$ and $\overline{XB}_i$ can obtain
masses via instanton effects. Also, they can couple to the singlets 
from the intersections of other seven-branes, and then obtain
masses from Higgs mechanism. For simplicity, in this paper, we will
assume that the masses for vector-like particles $XB_i$ and 
$\overline{XB}_i$ are around the $SU(5)\times U(1)_X$ unification scale.
In fact,  such kind of the flipped $SU(5)\times U(1)_X$ models 
without bulk vector-like particles $XB_i$ and $\overline{XB}_i$
can be constructed in F-theory as well, which will be presented 
elsewhere~\cite{JLNX-P}.

We consider three models: in Model I, we only introduce the vector-like
particles ($XF,~\overline{XF}$) and ($Xl,~\overline{Xl}$) with
universal mass $M_V$ around the TeV scale; in Model II,
 we  introduce the vector-like
particles ($XF,~\overline{XF}$) and ($Xl,~\overline{Xl}$) with
universal mass $M_V$, and the vector-like
particles ($Xl_{1},~\overline{Xl}_1$) and ($Xl_{2},~\overline{Xl}_2$)
with universal mass $M_{V'}$ around the $M_{23}$ scale;
in Model III,  we  introduce the vector-like
particles ($XF,~\overline{XF}$) with mass $M_V$, and the vector-like
particles ($Xl,~\overline{Xl}$) and ($Xl_{j},~\overline{Xl}_j$)
with  universal mass $M_{V'}$ for $j=1, 2,..., 6$. 
Thus, at the TeV scale, we have the $Z1$ set of vector-like
particles in Model I and Model II, and the $Z0$ set in Model III.
Also, the vector-like particles around $M_{23}$ scale can be 
considered  as  threshold corrections.

{\bf ${\cal F}$-$SU(5)$ from Free-Fermionic Constructions~--}~These
models have been studied before by three of us~\cite{Jiang:2006hf}. 
In Model IV and Model V, we introduce the $Z0$ and $Z1$ sets of 
vector-like particles, respectively. As in the 5/2 model in Table 4 
in Ref.~\cite{LNY}, there exist additional gauge symmetries
$SO(10)\times SU(4)\times U(1)^5$ in the hidden sector.
And there are five pairs of vector-like particles $XT_i$
and ${\overline{XT}}_i$ and one pair of vector-like particles ${ XT'}$
and ${\overline{XT'}}$~\cite{LNY} that transform non-trivially
under the SM gauge symmetry. These particles are only charged
under $SU(4) \times U(1)_X$, and their quantum numbers are
\begin{eqnarray}
&&  XT_i={\mathbf{(4,  {5\over 2})}}~,~
\overline{XT}_i={\mathbf{({\bar 4}, -{5\over 2})}} ~,~ \\
&& XT'={\mathbf{(4,  -{5\over 2})}}~,~
\overline{XT'}={\mathbf{({\bar 4}, {5\over 2})}} ~.~\,
\end{eqnarray}
Because these models essentially arise from weakly
coupled heterotic string theory,
we require the string-scale gauge coupling unification
as given in Eq.~(\ref{St-Unif}).

{\bf Gauge Coupling Unification~--}~Using the weak-scale
data in Ref.~\cite{Amsler:2008zz} and the renormalization
group equations in Ref.~\cite{Jiang:2006hf}, we study 
the gauge coupling 
unification at the two-loop level. For simplicity, we choose
$M_V=800 ~{\rm GeV} $ and the universal supesymmetry
breaking scale $M_S=800 ~{\rm GeV}$ in all of our models,
and choose $M_{V'}=1\times 10^{16}~{\rm GeV} $ 
in Models II and  III. 
In Models IV and V, $M_{V'}$ is determined from  
string-scale gauge coupling unification in 
Eq.~(\ref{St-Unif}).



\begin{table}[htb]
\begin{center}
\begin{tabular}{|c|c|c|c|c|c|}
\hline
Model   & $M_{23}$  &  $M_{V'}$ & $g_{23}$
& $g_{\rm U}$ & $M_{\rm U}$  \\
\hline
I  & $1.18 \times 10^{16}$  & $-$ & 1.193 & 1.165
&
$1.03 \times 10^{18}$   \\
II  & $1.18 \times 10^{16}$  & $10^{16}$ & 1.193 &
1.170 & $4.20 \times 10^{17}$   \\
III  & $1.18 \times 10^{16}$  & $10^{16}$ & 1.193
&
1.163 & $1.30 \times 10^{18}$   \\
IV  & $1.17 \times 10^{16}$  & $9.58 \times
10^{14}$ & 1.193 & 1.168 & $6.15 \times 10^{17}$   \\
V   & $1.18 \times 10^{16}$  & $2.98 \times
10^{17}$ & 1.193 &  1.168 & $6.16 \times 10^{17}$   \\
\hline
\end{tabular}
\end{center}
\caption{Mass scales in GeV unit and gauge couplings in the 
${\cal F}$-$SU(5)$ models for gauge coupling unification.}
\label{RGE-Data}
\end{table}



\begin{figure}[htb]
\centering
\includegraphics[width=9cm]{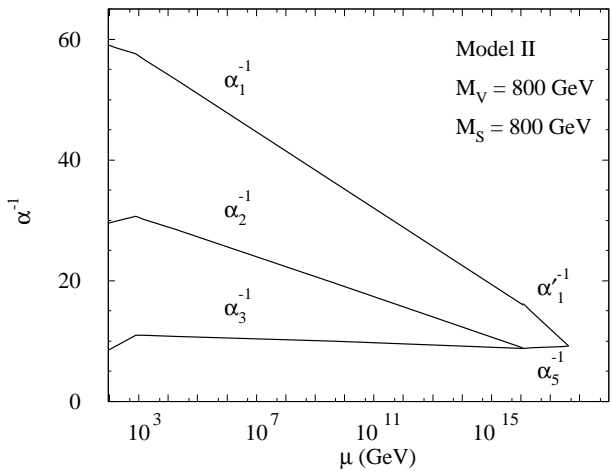}
\caption{Gauge coupling unification in 
Model II.}
\label{fig:2loopm}
\end{figure}

In Table~\ref{RGE-Data},  we present the masse scales $M_{23}$ and
$M_{U}$, and  the gauge couplings $g_{23}$ and $g_U$ in
our models, where $g_{23}$ is the $SU(2)_L\times SU(3)_C$ unified 
coupling and $g_U$ is the $SU(5)\times U(1)_X$ unified coupling.
As an example, we also plot the gauge coupling unification 
for Model II in Fig. \ref{fig:2loopm}.  From the Table~\ref{RGE-Data},
we obtain that $M_{23}$ is about $1.18\times 10^{16}~{\rm GeV}$
and $g_{23}$ is about $1.193$. In general, $g_{23}$ is smaller 
than the corresponding $g_U$ 
since $SU(5)$ is asymptotically free.
We emphasize that $g_{23}$ is stronger than that in the traditional
flipped $SU(5)\times U(1)_X$ models due to the TeV-scale
$Z0$ or $Z1$ set of vector-like particles, which
will be very important in the proton decay as discussed 
below.



{\bf Phenomenological Consequences~--}~Proton decay via 
dimension-5 operators from Higgsino exchange is suppressed.
Considering proton decay $p  \to e^+ \pi^0$ via dimension-6 
operator from heavy gauge boson exchange,
we obtain the proton life time~\cite{Ellis:2002vk}  
\begin{eqnarray}
\tau_p \simeq 9.97\times 10^{34} \left({{M_{23}}\over 
{1.18\times 10^{16}{\rm GeV}}}\right)^4
\left( {{1.193}\over {g_{23}}} \right)^4 ~{\rm yr}.~\,
\end{eqnarray}
Note that $M_{23}$ can be another factor 2/3 smaller 
due to  threshold corrections~\cite{Ellis:2002vk}, 
our models can definitely 
be tested at the future Hyper-Kamiokande proton decay experiment which
can search the proton life time via $p  \to e^+ \pi^0$ channel 
at least more than $10^{35}$ years~\cite{Nakamura:2003hk}. 
Also, the Kaluza-Klein modes of the gauge bosons could  further
enhance the proton decay. However, the details depend on the 
estimations of the bulk Green's functions for the gauge bosons
which have some unknown constants~\cite{Donagi:2008ca}.

Moreover, from Eq.~(\ref{potgut}), we obtain
that the neutrino masses and mixings can be explained via double seesaw 
mechanism~\cite{Ellis:1992nq}.
Also, the right-handed neutrino Majorana 
masses can be generated via the following dimension-5 operators after we 
integrate out the heavy Kaluza-Klein modes~\cite{Beasley:2008dc}
\begin{eqnarray}
W = {{y_{ij}^{\prime N}}\over {M_U}} F_i F_j \overline{H} \overline{H}~.~\,
\end{eqnarray}
So the neutrino masses and mixings can be generated via seesaw mechanism as well.
With leptogenesis~\cite{Fukugita:1986hr}, 
we can obtain the observed baryon asymmetry~\cite{Ellis:1992nq}.

In addition, from Eq.~(\ref{spgut}), we can naturally have 
the hybrid inflation where $\Phi$ is the inflaton
 field~\cite{Kyae:2005nv}. The inflation 
scale is related to the scale $M_{23}$. 
Because $M_{23}$ is at least one order smaller
than $M_U$, we solve the monopole problem.
Interestingly, we can 
generate the correct cosmic primordial density 
fluctuations~\cite{Spergel:2006hy}
\begin{eqnarray}
{{\delta \rho}\over {\rho}} \sim \left({{M_{23}}\over 
{g_{23} M_{\rm Pl}}}\right)^2 \sim  1.7\times 10^{-5}
~.~\,
\end{eqnarray}



{\bf Conclusions~--}~We have constructed three flipped $SU(5)\times U(1)_X$ 
models from F-theory, and considered two such models from free fermionic
constructions where we introduced vector-like particles at the TeV scale 
that can be produced at the LHC.
We studied gauge coupling unification, and achieved the decoupling 
scenario in F-theory models and the string-scale gauge coupling
unification in free fermionic models. Interestingly, 
proton decay is within the reach of the future Hyper-Kamiokande 
experiment.  In these models,  the neutrino masses 
and mixings can be explained via the double seesaw or seesaw mechanism, 
the observed baryon asymmetry can be obtained by leptogenesis, 
the hybrid inflation can be realized, the monopole problem can
be solved, and
the correct cosmic primodial density fluctuations can be generated.

{\bf Acknowledgments~--}~This research was supported in part by the
Cambridge-Mitchell Collaboration in Theoretical Cosmology (TL),
and by the DOE grant DE-FG03-95-Er-40917 (DVN).


\end{document}